%% file: main.tex
\documentclass[sigconf]{acmart}

\input{__packages}

\input{__macros}

\begin{document}

\title{Understanding Gaps in LLM Pipelines Towards Scalable Fuzzing Harness Generation: An Empirical Study and Enhancement}

\author{Kang Yang}
\affiliation{%
  \institution{University of Utah}
   \country{USA}
}

\author{Yunhang Zhang}
\affiliation{%
  \institution{University of Utah}
   \country{USA}
}

\author{Zichuan Li}
\affiliation{%
  \institution{University of Illinois Urbana-Champaign}
   \country{USA}
}

\author{Guanhong Tao}
\affiliation{%
  \institution{University of Utah}
   \country{USA}
}

\author{Jun Xu}
\affiliation{%
  \institution{University of Utah}
   \country{USA}
}

\author{Xiaojing Liao}
\affiliation{%
  \institution{University of Illinois Urbana-Champaign}
   \country{USA}
}
\input{sec_00_abstract}

\maketitle

\input{sec_01_introduction}
\input{sec_02_background}

\input{sec_03_study}

\input{sec_04_methodology}

\input{sec_05_evaluation}
\input{sec_06_relatedwork}

\input{sec_07_discussion}

\input{sec_09_conclusion}


\bibliographystyle{IEEEtran}
\bibliography{paper}
\input{sec_08_appendix}
\end{document}

%% file: __packages.tex
\usepackage[utf8]{inputenc}
\usepackage[most]{tcolorbox}
\usepackage{caption}
\usepackage{enumitem}
\setlist{nosep,leftmargin=8pt}
\usepackage{amsmath}
\usepackage{listings}
\usepackage{xcolor}
\usepackage[T1]{fontenc}
\usepackage{pifont}
\usepackage{booktabs}
\usepackage{multirow}
\usepackage{makecell}
\usepackage{minted}
\usepackage[normalem]{ulem}
\usepackage{hyperref}
\usepackage{courier}
\usepackage{wrapfig}

%% file: __macros.tex
\newcommand{\gptmini}{GPT5-Mini}

\newcommand{\qwen}{Qwen}

\newcommand{\claude}{Claude}
\newcommand{\ds}{DeepSeek}
\newcommand{\ossfuzzgen}{OSS-Fuzz-Gen}
\newcommand{\ossfuzz}{OSS-Fuzz}
\newcommand{\issta}{LLM4FDG}
\newcommand{\sherpa}{Sherpa}
\newcommand{\our}{HarnessAgent}
\tcbset{
  mypromptbox/.style={
    enhanced,
    colback=gray!5,
    colframe=black!70,
    fonttitle=\bfseries,
    title={#1},
    boxrule=0.4pt,
    arc=3pt,
    left=6pt,
    right=6pt,
    top=4pt,
    bottom=4pt
  }
}

\setminted{
  fontsize=\scriptsize,
  baselinestretch=1.0,
  fontfamily=tt,
  frame=single,
  framesep=2mm,
  bgcolor=white,
  linenos,
  numbersep=6pt,
  breaklines=true,
}

\definecolor{codegray}{rgb}{0.5,0.5,0.5}
\definecolor{codeblue}{rgb}{0.0,0.0,0.6}
\definecolor{codegreen}{rgb}{0,0.5,0}
\definecolor{backgray}{rgb}{0.97,0.97,0.97}

\lstdefinestyle{cpp}{
  language=C,
  basicstyle=\ttfamily\footnotesize,
  commentstyle=\color{codegreen}\itshape,
  keywordstyle=\color{codeblue}\bfseries,
  stringstyle=\color{red!70!black},
  numbers=left,
  numberstyle=\tiny\color{codegray},
  numbersep=6pt,
  stepnumber=1,
  tabsize=2,
  showstringspaces=false,
  breaklines=true,
  frame=single,
  framerule=0.3pt,
  rulecolor=\color{black!20},
  captionpos=b,
  xleftmargin=1em,
  framexleftmargin=1.2em,
  morekeywords={
    FILE, size_t, uint8_t, uint16_t, uint32_t, uint64_t,
    int8_t, int16_t, int32_t, int64_t,
    struct, constexpr, nullptr, override, final,
    const, void, char, bool, unsigned, long, short,
    inline, static, enum, typedef
  }
}

\colorlet{mgray}{gray!90!black}
\definecolor{cadmiumgreen}{rgb}{0.0, 0.42, 0.24}

\lstdefinestyle{c}{
    language=C++,
	frame			= tb,
	aboveskip		= 2mm,
	belowskip		= 2mm,
	showstringspaces = false,
	showtabs		= false, 
	columns			= flexible,
    basicstyle=\fontencoding{T1}\scriptsize\fontfamily{lmtt}\selectfont,
	numbers			= left,
	stepnumber		= 1, 
	numbersep		= 0.5em, 
	numberstyle		= \scriptsize\color{cyan},
	keywordstyle	= \color{cadmiumgreen},
	commentstyle	= \color{mgray},
	stringstyle		= \color{violet},
	breaklines		= true,
	breakatwhitespace = true,
	tabsize			= 2,
	captionpos		= b, 
	breaklines		= true, 
	breakatwhitespace = true, 
    morekeywords={
    FILE, size_t, uint8_t, uint16_t, uint32_t, uint64_t,
    int8_t, int16_t, int32_t, int64_t,
    struct, constexpr, nullptr, override, final,
    const, void, char, bool, unsigned, long, short,
    inline, static, enum, typedef, isc_buffer_t,isc_result_t,dns_message_t,
    isc_mem_t,
  },
  emph={
    LLVMFuzzerTestOneInput,
    isc_mem_create,
    isc_buffer_create_mock,
    dns_message_create,
    dns_message_parse,
    dns_message_destroy,
    isc_mem_destroy,
    dns_message_detach,
    isc_mem_detach,
    isc_buffer_init
  },
  emphstyle=\color{blue},
  emph=[2]{NULL,0, DNS_MESSAGE_INTENTPARSE, ISC_R_SUCCESS},
  emphstyle=[2]{\color{violet}}
}

\lstdefinestyle{cbig}{
    language=C++,
	frame			= tb,
	aboveskip		= 2mm,
	belowskip		= 2mm,
	showstringspaces = false,
	showtabs		= false, 
	columns			= flexible,
    basicstyle=\fontencoding{T1}\footnotesize\fontfamily{lmtt}\selectfont,
	numbers			= left,
	stepnumber		= 1, 
	numbersep		= 0.5em, 
	numberstyle		= \scriptsize\color{cyan},
	keywordstyle	= \color{cadmiumgreen},
	commentstyle	= \color{mgray},
	stringstyle		= \color{violet},
	breaklines		= true,
	breakatwhitespace = true,
	tabsize			= 2,
	captionpos		= b, 
	breaklines		= true, 
	breakatwhitespace = true, 
    morekeywords={
    FILE, size_t, uint8_t, uint16_t, uint32_t, uint64_t,
    int8_t, int16_t, int32_t, int64_t,
    struct, constexpr, nullptr, override, final,
    const, void, char, bool, unsigned, long, short,
    inline, static, enum, typedef, isc_buffer_t,isc_result_t,dns_message_t,
    isc_mem_t,
  },
  emph={
    LLVMFuzzerTestOneInput,
    isc_mem_create,
    isc_buffer_create_mock,
    dns_message_create,
    dns_message_parse,
    dns_message_destroy,
    isc_mem_destroy,
    dns_message_detach,
    isc_mem_detach,
    isc_buffer_init
  },
  emphstyle=\color{blue},
  emph=[2]{NULL,0, DNS_MESSAGE_INTENTPARSE, ISC_R_SUCCESS},
  emphstyle=[2]{\color{violet}}
}

\newtcolorbox{highlightbox}{
  colback=gray!15,        
  colframe=gray!15,       
  arc=2mm,                
  boxrule=0pt,            
  left=1.5mm, right=1.5mm,
  top=0.5mm, bottom=0.5mm,
  enlarge top by=0mm,     
  enlarge bottom by=0mm,  
}


%% file: sec_00_abstract.tex
\begin{abstract}

Large language model (LLM)-based techniques have achieved notable progress in fuzz harness generation. However, applying them to arbitrary functions \textit{at scale} remains difficult---generated harnesses often fail to compile or, worse, compile but remain logically ineffective. What factors drive success and what limitations hinder current methods remain unclear.

To answer these questions, we conduct an empirical study on state-of-the-art LLM-based harness generation frameworks across 29 OSS-Fuzz projects. Our study establishes two pillars of success: contextual information to guide generation and pre-structured pipelines that ensure workflow stability. However, significant gaps remain: (1) existing context retrieval methods lack the robustness to reliably obtain necessary information across diverse projects; (2) current validation mechanisms fail to detect logically ineffective harnesses, such as those containing fabricated definitions; and (3) compilation workflows are brittle, unable to distinguish harness-level errors from build-configuration issues. We demonstrate the utility of these findings by enhancing existing techniques with a hybrid tool pool for robust context retrieval, an enhanced validation pipeline, and a compilation-error triage strategy. Evaluated on 243 OSS-Fuzz projects (65 C and 178 C++), the enhanced approach improves the three-shot success rate by approximately 20\% over state-of-the-art techniques, reaching 87\% for C and 81\% for C++. Our one-hour fuzzing results show that more than 75\% of the generated harnesses increase target-function coverage, surpassing baselines by over 10\%. In addition, the enhanced approach identified 15 new vulnerabilities across 11 real-world projects.
\end{abstract}

%% file: sec_01_introduction.tex
\section{Introduction}

Fuzz testing, which continuously generates and mutates inputs for executable target programs to trigger unexpected behaviors, has become the de facto standard for discovering zero-day vulnerabilities in modern software systems~\cite{fuzzing_roadmap, bohme2016coverage, AFL, libfuzzer}.
%
%
However, the effectiveness of fuzzing heavily depends on the quality of its \textit{harness} (or fuzz driver), which is a small yet crucial piece of code that receives mutated inputs and invokes the target program. 
The correctness and robustness of the harness are vital: improper function usage or incomplete state handling can lead to extensive false positives or negatives, wasting testing resources and demanding additional manual validation. 

This challenge becomes even more pronounced when targeting arbitrary functions across large and diverse software projects. Many functions rely on implicit dependencies, project-specific data structures, or complex initialization procedures that are rarely documented. Constructing a valid harness requires inferring the correct calling context, preparing appropriate inputs, and understanding the surrounding environment---tasks that demand deep program understanding and significant manual effort~\cite{issta, OSS-Fuzz-Gen}. 
As a result, fuzzing arbitrary functions at scale remains difficult, limiting the reach of fuzzing to only a subset of potentially vulnerable code regions.

Meanwhile, the research community has increasingly investigated automatic harness generation techniques, which aim to automatically infer valid and effective input structures, initialization routines, and/or call sequences. Traditional approaches through program analysis~\cite{FUDGE, apicraft,jung2021winnie} or
runtime feedback~\cite{afgen, hopper, green2022graphfuzz} often struggle with incorrect execution contexts or function invocations, resulting in a low success rate.
Recent advances in Large Language Models (LLMs) have shown promise in automating fuzz harness generation~\cite{issta, OSS-Fuzz-Gen, sherpa2024, deng2023large, deng2024large}, successfully producing valid harnesses for many target functions. However, scaling these techniques to arbitrary functions across diverse projects remains difficult based on our observations, and the factors that determine success or failure are not well understood.

\vspace{3pt}
\noindent \textbf{Our Study.}
To better understand these issues, we conduct an empirical study on state-of-the-art LLM-based harness generation frameworks across 29 OSS-Fuzz projects. We examine three representative systems---\issta~\cite{issta}, \ossfuzzgen~\cite{OSS-Fuzz-Gen}, and \sherpa~\cite{sherpa2024}---and systematically analyze their four-step workflow: generation, compilation, fixing, and validation.

Our analysis confirms two effective practices. First, providing contextual information---such as header files and function usage examples---significantly improves harness validity; among all context types, function usage delivers the most consistent improvement across models. Second, pre-structured pipelines with carefully engineered workflows substantially outperform LLM-managed approaches, which struggle with build errors and lack the reliability needed for large-scale deployment.

Despite these insights, significant gaps remain in current systems. First, existing context retrieval methods are not robust enough to reliably obtain necessary information---such as symbol definitions and header files---across diverse C/C++ projects with varying conventions and build configurations. Second, current validation mechanisms fail to detect logically ineffective harnesses; we observe that LLMs often fabricate fake function definitions to bypass compilation checks (e.g., over 10 out of 56 harnesses in our study contain such fabrications). Third, compilation workflows are brittle: they cannot distinguish harness-level errors from build-configuration issues, causing models to repeatedly generate fixes misaligned with actual build feedback.

\vspace{3pt}
\noindent \textbf{Enhancing with Findings.}
We demonstrate the utility of these findings by enhancing existing harness generation techniques. Specifically, we address each identified gap with a targeted solution.

First, to address the context retrieval gap, we introduce a hybrid tool pool combining a Language Server Protocol (LSP~\cite{lsp}) interface for precise symbol queries with a grammar-tree parser (Tree-sitter~\cite{treesitter}) for robust pattern matching. Second, to counter LLM validation exploitation, we implement structural checks that parse generated harnesses and verify they don't contain fabricated function definitions. Third, to handle brittle compilation workflows, we add a compilation-error triage logic that classifies build failures and routes them to appropriate fix actions, preventing misaligned corrections.

We evaluate the enhanced approach on 243 target functions from OSS-Fuzz projects (65 C and 178 C++). The results show that these enhancements improve the three-shot success rate by approximately 20\% over state-of-the-art baselines, reaching 87\% for C and 81\% for C++. In one-hour fuzzing experiments, over 78\% of the generated harnesses led to measurable increases in target-function coverage, exceeding baselines by more than 10\%. Furthermore, the enhanced approach uncovered 15 previously unknown vulnerabilities across 11 real-world projects.

%

\vspace{3pt}
\noindent \textbf{Contributions.} We summarize the contributions as follows:

\noindent $\bullet$ We conduct an empirical study on state-of-the-art LLM-based harness generation frameworks across 29 OSS-Fuzz C projects. Our analysis identifies two pillars of success---contextual information and pre-structured pipelines---and three critical gaps in current systems.

\noindent $\bullet$ We demonstrate the utility of these findings by enhancing existing techniques with a hybrid context retrieval tool pool, structural validation checks, and compilation-error triage logic.

\noindent $\bullet$ We evaluate the enhanced approach on 243 target functions from OSS-Fuzz projects (65 C and 178 C++). The results show a 20\% improvement in success rate over state-of-the-art baselines, and the approach uncovered 15 previously unknown vulnerabilities. We release the datasets and source code at \url{https://anonymous.4open.science/r/LLM-reasoning-agents-5B93/}.\looseness=-1



%% file: sec_02_background.tex
\section{Background}

\subsection{Harnesses for Program Fuzzing}

Fuzzing~\cite{fuzzing_roadmap, bohme2016coverage, AFL, libfuzzer} tests software with invalid, unexpected, or random inputs, but deep code components may not accept such inputs directly. A \textit{harness}~\cite{fuzzgen} bridges the \textit{fuzzer} and the \textit{target program} by translating inputs into target interfaces, setting up required state, and handling initialization/teardown; it may also add basic error handling and isolation so failures are reported without stopping fuzzing.
Harnesses are built manually~\cite{srlabsFuzzingMade} or via automated program-analysis methods~\cite{fuzzgen, FUDGE, capiharness, rahalkar2023automated, jung2021winnie, green2022graphfuzz, rubick, utopia, IntelliGen}. Manual construction is tedious and time-consuming, requiring domain knowledge and project understanding, while automated approaches use analysis or runtime feedback to infer dependencies but often fail due to incorrect contexts or invocations, leading to low success rates.\looseness=-1

\subsection{LLM for Harness Generation\label{sec:fuzz_workflow}}
\input{figure/fuzzframework}

To overcome the limitations of manual construction and traditional program-analysis approaches, recent work has explored using Large Language Models (LLMs) for harness generation~\cite{promptfuzzingccs24, promefuzzccs25, ckgfuzzer, issta, OSS-Fuzz-Gen}. LLMs can leverage their code understanding capabilities to synthesize harnesses without requiring complex static analysis or runtime instrumentation.

\autoref{fig:fuzzframework} summarizes the unified LLM-based workflow from prior work~\cite{issta, OSS-Fuzz-Gen}, consisting of \textit{generation}, \textit{compilation}, \textit{fixing}, and \textit{validation}. The LLM first generates a harness from provided information (\ding{182}) and compiles it with the target into a fuzz target (\ding{183}); compilation failures trigger fixing using error messages (\ding{184}). If compilation succeeds, validation checks criteria such as target invocation, coverage increase, and crashes (\ding{185}); validation errors are fed back to fixing (\ding{186}). This loop repeats until validation succeeds or a preset iteration limit is reached.

\smallskip
\noindent
\textbf{Contextual Information.}
Existing methods provide contextual information to the LLM during both generation and fixing~\cite{contextenginner}. For generation, context typically includes the target function's definition, usage examples, and related headers. For fixing, context relates to the specific compilation or validation error. \issta~\cite{issta} supplies the declaration and usage of the first function appearing in the error message (\textit{function-based context}), while \ossfuzzgen~\cite{OSS-Fuzz-Gen} defines regular patterns to match specific errors and provide corresponding code information (\textit{rule-based context}).

\smallskip
\noindent
\textbf{Validation Criteria.}
Three criteria are commonly used to validate harness correctness~\cite{issta, OSS-Fuzz-Gen}: (1) \textit{Call check}: ensures the target function is invoked; (2) \textit{Fuzzing check}: runs the harness briefly (e.g., 30--60 seconds) to detect crashes caused by harness errors rather than target bugs; and (3) \textit{Code coverage check}: verifies that fuzzing increases coverage, indicating the harness effectively exercises the target function.

%% file: figure/fuzzframework.tex
\begin{figure}[t]
    \centering
    \includegraphics[width=\columnwidth]{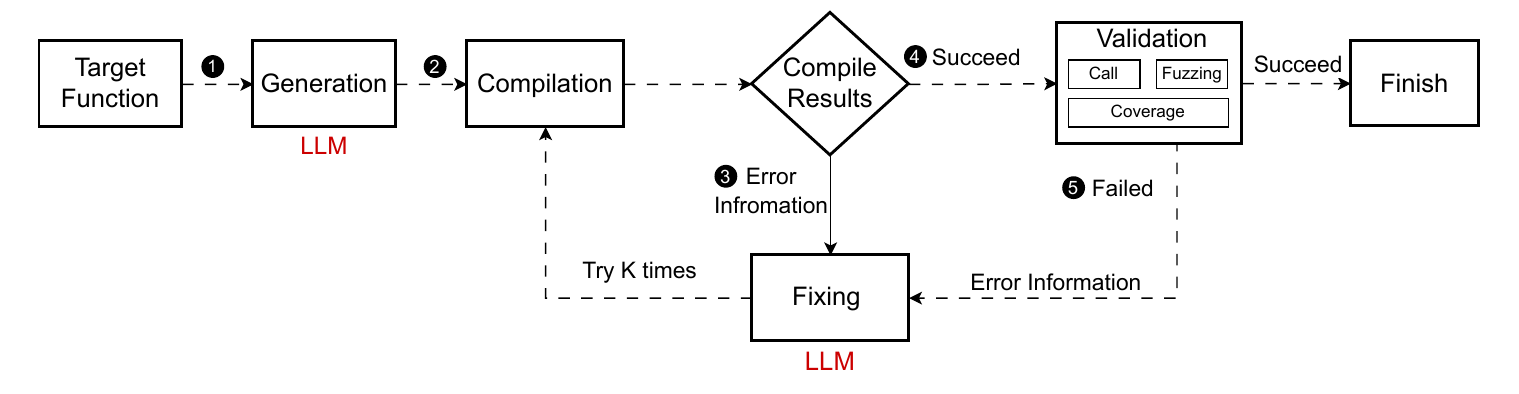}
    \vspace{-2.5em}
    \caption{The General Harness Generation Workflow with LLMs.}
    \label{fig:fuzzframework}
\end{figure}

%% file: sec_03_study.tex
\section{Study Setup}
\label{sec:study}

To understand the status quo of LLM-based harness generation, including success factors and remaining obstacles, we conduct an empirical study on state-of-the-art frameworks. Below we describe the study setup.

\subsection{Dataset}
\label{subsec:datasets}
We adopt the dataset used in \issta~\cite{issta}, which comprises 86 functions across 30 OSS-Fuzz C projects. The spdk~\cite{spdk} project failed to build fuzzers, so we excluded it from our study, resulting in a final dataset of 85 functions. 

\subsection{Generation Frameworks Under Study}
\label{subsec:baselines}
We explore three state-of-the-art approaches in automatic harness generation.

\begin{itemize}[leftmargin=12pt]
    \item \textbf{\issta}~\cite{issta} leverages LLMs to automatically generate harnesses with a focus on the generation module. It designs six prompting strategies and explores how different prompts impact the effectiveness. 

    \item \textbf{\ossfuzzgen}~\cite{OSS-Fuzz-Gen} is a framework that generates fuzz targets for real-world C/C++/Java/Python projects with various LLMs. It adopts Fuzz Introspector~\cite{fuzz-introspector} to provide source code information, such as header files and definitions. When fixing harnesses, it uses a rule-based approach to provide corresponding code information to the LLM.

    \item \textbf{\sherpa}~\cite{sherpa2024} is an LLM pipeline that automatically identifies high-value, attacker-controlled entry points in OSS-Fuzz projects and generates corresponding fuzz harnesses. In detail, it adopts the Codex~\cite{codex} agent to perform the entire harness generation, compilation, and fixing loop.
\end{itemize}

Note that \issta{} and \ossfuzzgen{} focus on generating a harness for a specific target function leveraging LLMs. They both follow the workflow described in~\S\ref{sec:fuzz_workflow}, and the only difference lies in the provided context in the generation and fixing modules.

\subsection{LLMs}
We considered four foundation models with different capabilities and design origins to examine their impact on harness-generation performance: Claude Haiku 4.5~\cite{claudehaiku}, GPT-5-Mini~\cite{gpt5}, Qwen3-Coder~\cite{qwen3}, and DeepSeek V3.2~\cite{deepseek}. Throughout the rest of the paper, we refer to them using the shorter names \claude, \gptmini, \qwen, and \ds~for simplicity. Among them, \qwen~ and \ds~ are open-source models, while the other two are closed-source models. We set the temperature to 1 for a better tradeoff between randomness and creativity of the model's output, which is also the default temperature for most LLMs.

\input{table/context}

\subsection{Reasoning Contexts}
\autoref{table:context} summarizes the contexts used by existing techniques.
These contexts fall into five categories: \textit{the header file}, \textit{the function definition}, \textit{the function usage}, \textit{the function document}, and \textit{the harness examples}. Note that due to the lack of documentation for most of the functions, we exclude this context from our experiment.  Regarding the contextual information for harness fixing, we consider the \textit{rule-based context} and \textit{function-based context} from \ossfuzzgen~and \issta~mentioned in~\S\ref{sec:fuzz_workflow}, respectively.
For implementation, we use the official codebase of \ossfuzzgen~\cite{OSS-Fuzz-Gen} without modification. For \issta, we re-implement the approach following the methodology described in the original paper~\cite{issta} to ensure consistency and fair comparison.

We also introduce a \textbf{Minimal-Context Baseline} to isolate the effect of contextual information. This baseline follows \issta{} and \ossfuzzgen{} (\S\ref{sec:fuzz_workflow}) but provides only the task description and target-function declaration for generation. During fixing, the model receives the harness code plus a minimal error summary: pattern-mined compiler errors, a short message for failed validation (e.g., ``\textit{the target function is not called}''), and crash reason/stack for fuzzing failures.

\subsection{Validation Criteria}
\label{subsec:criteria}
By default, if the harness passes the following validation criteria, we consider it as successful.

$\bullet$ The harness does not contain a fake definition for the target function, and the target function is called.

$\bullet$ The harness must be compiled and fuzzed for 60 seconds without any crashes.

$\bullet$ The code coverage increases during the fuzzing.

$\bullet$ The harness also needs to pass the semantic testing established in \issta~\cite{issta}.

Each function undergoes three harness generation attempts, and we consider it successful if at least one of these attempts passes the validation module. The \textit{success number} across the entire dataset serves as our performance metric.


\section{Results and Findings}
\label{sec:study_results}

Our study reveals both promising practices and significant gaps in current LLM-based harness generation. We organize our findings into three categories: what works well (\textit{The Good}), what fails (\textit{The Bad}), and what breaks at scale (\textit{The Ugly}).

\subsection{The Good}

\subsubsection{Context Provision Works}

Contextual information significantly improves harness generation success rates. To quantify this effect, we begin with minimal context and progressively incorporate additional information to assess their individual and combined impact.

As shown in \autoref{table:ablation}, compared to the \emph{Minimal-Context Baseline}, adding project-specific contexts improves the success rate across all models. Starting from the baseline, the number of successful functions is modest (e.g., 49 for \qwen), but supplying even lightweight context improves the results (e.g., from 49 to 61 for \qwen).\looseness=-1

\input{table/ablation}

\textit{Not all contexts contribute equally.} \emph{Header files} improve compilation success by resolving symbol dependencies, but their contribution is inconsistent---for example, headers do not improve performance for \claude{} (50 to 49) but boost \qwen{} (49 to 58). By contrast, \emph{Function Usage} delivers semantic priors like argument initialization and call sequences, producing the strongest and most uniform improvement: moving from baseline to \emph{Function Usage} yields 15 more successful cases for \claude{} (49 to 64) and around 12 cases on average across models. \emph{Function Definition} and \emph{Harness Example} contribute only marginal gains, as they offer limited guidance specific to the target function's intended behavior.

\begin{highlightbox}
\textbf{Takeaway.} Existing methods effectively leverage contextual information to improve harness generation, with success rate gains of up to 20\%.
\end{highlightbox}

\subsubsection{Pre-Structured Pipelines Work}

To understand which workflow schema is more effective, we compare \sherpa~\cite{sherpa2024} (an LLM-managed planner)  with \issta~\cite{issta} and \ossfuzzgen~\cite{OSS-Fuzz-Gen} (pre-structured systems) on the same dataset. As shown in \autoref{fig:planning}, \sherpa{} produces only 9 successful harnesses, far less than the pre-structured systems \ossfuzzgen{} and \issta.
This gap is mainly due to failing to generate new fuzz targets (\sherpa{} creates a new fuzzer, but the agent reuses the original fuzzer) and frequent build errors; moreover, \sherpa{} lacks an explicit compilation workflow and relies on the agent, which often fails to fix build errors within retry limits. In contrast, pre-structured systems use carefully engineered workflows that improve stability.\looseness=-1

In addition to the ineffectiveness, the LLM-managed planner also suffers from efficiency issues. For example, \sherpa{} does not provide concrete function usage examples or integration guidelines; as a result, the agent often spends substantial time walking through the project to understand and locate the necessary dependencies.
Although the LLM-managed planner offers simplicity and flexibility, it struggles to fix build errors reliably and efficiently.

\begin{highlightbox}
\textbf{Takeaway.} Pre-structured pipelines in existing methods provide workflow stability, achieving higher success rates than LLM-managed alternatives.
\end{highlightbox}

\input{figure/study_planning}

\subsection{The Bad}
\label{subsec:limitation}

Despite the above successes, current methods exhibit notable limitations in context retrieval and validation.

\subsubsection{Flawed Context Retrieval}

Current context retrieval methods suffer from two issues: incomplete coverage and passive delivery.

\smallskip\noindent
\textbf{Incomplete Context:}
Existing methods provide insufficient context. In the generation phase, while they consider various contexts, they still miss critical pieces. \autoref{lst:driver} illustrates a harness for \lstinline[style=cbig]|dns_message_parse|, generated by LLM. Constructing this harness requires initializing an \lstinline[style=cbig]|isc_buffer_t| object using  \lstinline[style=cbig]|isc_buffer_init| API (line \texttt{11}), before passing it to the target function (line \texttt{12}). The existing methods fail to provide the initialization context, leading the LLM to rely on its pretrained knowledge. As a result, the LLM fabricates an initialization function \lstinline[style=cbig]|'create_buffer_mock'| instead of using the correct API, failing to generate a valid harness. 

The situation of missing context is even worse in the fixing stage. \issta{} only provides function information and ignores class or struct errors; \ossfuzzgen{} pre-defines seven rules but misses many other error types. As a result, the fixing stage shows minimal improvement across all models (\autoref{table:ablation}).

\input{table/bind9_list}

\smallskip\noindent
\textbf{Passive Context Provision:}
Current methods collect a pre-defined set of contexts before generation, and retrieve additional contexts based on error messages when compilation fails. This reactive approach does not always work.

In the above example (\autoref{lst:driver}), besides faking the initialization function, the LLM also calls two deprecated APIs (\lstinline[style=cbig]|isc_mem_destroy| and \lstinline[style=cbig]| dns_message_destroy|) for cleanup. These lead to the following compilation errors:\looseness=-1
\begin{itemize}[leftmargin=2em, itemsep=-2pt, parsep=0pt, partopsep=0pt]
    \item \lstinline[style=c]|error: no previous prototype for function 'create_buffer_mock'|
    \item \lstinline[style=c]|warning: call to undeclared function 'dns_message_destroy'|
    \item \lstinline[style=c]|warning: call to undeclared function 'isc_mem_destroy'|
\end{itemize}

To fix such errors, previous passive approaches attempt to retrieve function information from the source code; however, no valid information is available because the erroneous functions are either deprecated or nonexistent. This motivates equipping LLMs with tools to proactively query contexts during generation and fixing, rather than relying on pre-collected or error-triggered retrieval.

\input{table/ablation_fake}

\subsubsection{Insufficient Validation}

When unable to find correct APIs, the model generates placeholder or fake functions that compile successfully but provide no actual fuzzing value. Existing validation methods fail to detect such fabrications.

As shown in \autoref{table:ablation_fake}, without additional checks, the apparent success rate is inflated. For instance, \gptmini~achieves 56 passes under \textit{Minimal} context, but 10 of these contain fake definitions, leaving only 46 genuine harnesses. Under \textit{Header+Function Usage}, 5 out of 70 passes are fake. This trend holds across models.

To validate that this is indeed an insufficient validation problem, we add a \textit{fake check} using \textit{Tree-Sitter} to detect fabricated definitions. With this check enabled, \gptmini achieves 49 passes, exceeding the 46 genuine harnesses obtained without the check, as the validation encourages the model to continue generating until a valid harness is produced. More context suppresses fake generation (10 fakes under \textit{Minimal} vs. 5 under \textit{Header+Function Usage}), but the check remains necessary to ensure reliability.

\begin{highlightbox}
\textbf{Takeaway.} Existing methods fail to proactively deliver the complete contextual information, limiting the effectiveness and reliability of harness generation. Furthermore, insufficient context encourages the LLM to generate fake definitions to bypass validation.
\end{highlightbox}

\subsection{The Ugly}

Beyond algorithmic limitations, scaling to real-world projects exposes engineering challenges in the compilation pipeline—issues often overlooked in research but critical for practical deployment.

While \ossfuzz{} provides a universal build script for project compilation, it is tailored to existing harnesses. When compiling a newly generated harness with this script, the process often raises compilation errors due to discrepancies in header inclusions, linking configurations, and other build dependencies. While small-scale studies can rely on manual fixes, scaling to hundreds of projects demands an automated and fault-tolerant compilation pipeline. The framework must be capable of distinguishing between compilation errors originating from the build script and those from the generated harness code, and resolving them autonomously. Achieving this requires fine-grained error analysis and specific strategies to mitigate compilation failures effectively.

Together, these challenges underscore that scaling LLM-based harness generation is no longer a matter of model capability alone. It requires a coordinated system-level effort to achieve automation across the entire pipeline.

%% file: table/context.tex
\begin{table}[t]
    \centering
    \caption{Contexts used by existing methods~\cite{issta}\cite{OSS-Fuzz-Gen} for harness generation and fixing. \textit{TF} and \textit{EF} represent ``target function'' and ``error function'', respectively.}
    \label{table:context}
    \footnotesize
    \tabcolsep=4.1pt
    \begin{tabular}{ccll}
        \toprule
        \textbf{Method} & \textbf{Generation}  & \makecell{\textbf{Fixing for} \\ \textbf{Compile Error}}    & \makecell{\textbf{Fixing for} \\ \textbf{Validation Error}}    \\
        \midrule
        \makecell{\textbf{\issta}~\cite{issta}}  &\makecell{Header File \\ TF Usage\\ TF Document\\ TF  Definition\\ TF Declaration} & \makecell{Error Info\\ EF Usage\\ EF Declaration} & \makecell{Crash Info \\ EF Usage\\ EF Declaration} \\
        \midrule
        \makecell{\textbf{\ossfuzzgen}~\cite{OSS-Fuzz-Gen}}    & \makecell{Header File \\ TF Usage\\ TF  Definition\\ TF Declaration \\ Harness Example }           & \makecell{Error Info.\\ Rule-Based Info.}                                   & \makecell{Crash Info.}      \\
        \bottomrule
    \end{tabular}
\end{table}



%% file: table/ablation.tex
\begin{table}[!t]
  \centering
  \caption{Harness generation performance when different contexts are used. As missing headers can invalidate other contextual elements, we combine all remaining context with header information.\label{table:ablation}}
  \tabcolsep=0.6pt
  \footnotesize
  \begin{tabular}{clcccc}
  \toprule
  \textbf{Module} & \textbf{Context}                         &\textbf{Claude} &\textbf{GPT5-Mini}   & \textbf{Qwen} & \textbf{DeepSeek}\\
  \midrule
  - &  Minimal-Context Baseline & 50     & 49       &  49    &  42  \\
  \midrule
  \multirow{6}{*}{\makecell{\textbf{Generation}}}     & Header File
          &           49  & 55       &  58     &  48  \\
      \cmidrule(l){2-6}
      &  Header File + Func. Usage         & 64  & 67   &  61    & 57    \\
      \cmidrule(l){2-6}
    & Header File + Func. Definition        & 55    &  54    & 60    &  53   \\
        \cmidrule(l){2-6}
    &  Header File + Harness Example     & 54  &    56    & 61   &  52     \\
    \midrule

  \multirow{2.5}{*}{\textbf{Fixing}} 
      & Header File + \issta             & 56     & 54      & 58     & 52  \\
      \cmidrule(lr){2-6}
      &    Header File + \ossfuzzgen & 58    &  55      & 56    & 50   \\

  \bottomrule
  \end{tabular}
\end{table}

%% file: figure/study_planning.tex
\begin{figure}[t!]
    \centering
    \includegraphics[width=1\columnwidth]{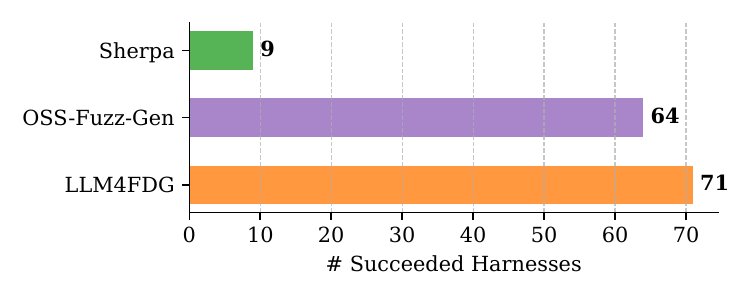}
    \vspace{-2em}
    \caption{Number of successful harnesses generated by existing methods.}
    \label{fig:planning}
\end{figure}


%% file: table/bind9_list.tex
\begin{listing}[t]
\caption{Example harness for function \lstinline[style=cbig]!'dns_message_parse'! from project Bind9~\cite{bind9}, generated by LLM.}
\label{lst:driver}
\begin{lstlisting}[style=c]
#include "dns/message.h"
#include ...
// Fake implementation of buffer creation
isc_buffer_t* create_buffer_mock(const uint8_t *data, size_t size) {
    isc_buffer_t *buffer = (isc_buffer_t*)malloc(sizeof(isc_buffer_t));
    if (buffer) {     buffer->data = data; ... }     // No such field
    return buffer;
}
int LLVMFuzzerTestOneInput(const uint8_t *Data, size_t Size) {
    ...
    isc_buffer_t *buffer = create_buffer_mock(Data, Size); // Fake function to init `buffer`
    result = dns_message_parse(..., &buffer, 0);  // Parse the DNS message using fuzz input
    dns_message_destroy(...);     // Clean up
    isc_mem_destroy(...);
    return 0;
}
\end{lstlisting}
\end{listing}

%% file: table/ablation_fake.tex
\begin{table}[t!]
\centering
\footnotesize
\caption{Harness generation results under different contexts (\textit{Minimal-Context Baseline } and \textit{Header+Function Usage}) with and without fake-check validation. ``\texttt{\#Pass}'' represents the total number of harnesses passing validation, while ``\texttt{\#Fake}'' indicates the number of harnesses with fake definitions among ``\texttt{\#Pass}'' cases.\label{table:ablation_fake}}
\tabcolsep=1.6pt
\begin{tabular}{l|cccc|cccc}
\toprule
\multirow{3}{*}{\textbf{Model}} & \multicolumn{4}{c|}{\textbf{Minimal-Context Baseline}}   & \multicolumn{4}{c}{\textbf{Header+Function Usage}} \\
\cmidrule(lr){2-5} \cmidrule(lr){6-9}
    & \multicolumn{2}{c}{With Check} & \multicolumn{2}{c}{W/O Check} & \multicolumn{2}{|c}{With Check} & \multicolumn{2}{c}{W/O Check} \\
     \cmidrule(lr){2-3} \cmidrule(lr){4-5}  \cmidrule(lr){6-7} \cmidrule(lr){8-9}
    & \texttt{\#Pass} & \texttt{\#Fake}  & \texttt{\#Pass}  & \texttt{\#Fake}   & \texttt{\#Pass}  & \texttt{\#Fake}  & \texttt{\#Pass}  & \texttt{\#Fake}  \\

\midrule
\textbf{\claude}         & 50     & 0    & 56       & 6       & 64     &  0       & 64     & 0   \\

\textbf{\gptmini}       & 49    & 0     & 56       & 10       & 67     & 0      & 70     &  5 \\

\textbf{\ds}            & 42    & 0    &  61      &  24      &  57    &  0      & 70     &  23  \\    

\textbf{\qwen}          & 49   & 0    &  57      &  13      &  61    &  0       & 64     & 3    \\    
\bottomrule
\end{tabular}
\end{table}






%% file: sec_04_methodology.tex
\input{figure/studyframework}
\section{Enhancing Harness Generation}

\subsection{Overview}

To demonstrate the utility of our findings, we enhance existing harness generation techniques with targeted solutions for each identified barrier. \autoref{fig:studyframework} illustrates the enhanced framework, which we refer to as \our, augmenting the standard LLM-based workflow (\S\ref{sec:fuzz_workflow}) with three components: (1) a \textit{tool pool} for proactive context retrieval (\S\ref{subsec:toolpool}), (2) an \textit{enhanced validation} module to detect fabricated definitions (\S\ref{subsec:validation}), and (3) a \textit{compilation-error triage} strategy for robust build integration (\S\ref{subsec:compilation}). Together, these enhancements address the barriers identified in \S\ref{sec:study_results} while preserving the structured pipeline that our study found effective.




\subsection{Tool-Augmented Context Retrieval}
\label{subsec:toolpool}

Guided by the insights derived from~\S\ref{sec:study}, our principle is to provide the LLM with only minimal and essential context to reduce noise and unrelated information, while equipping it with a rich set of tools to retrieve and refine the information it truly needs. 



\subsubsection{Tool Pool}

To address the incomplete and passive context provision identified in our study, we equip the LLM with a set of tools for proactive code retrieval. These tools leverage two complementary backends, a \textit{Language Server Protocol (LSP)}~\cite{lsp} interface and a \textit{grammar tree parser}, to reliably obtain symbol source code and header files across diverse projects. We expose seven tools in four categories:

\begin{itemize}[leftmargin=8pt, itemsep=2pt]
    \item \textit{Symbol Source Code Tools.} Retrieve symbol definitions, declarations, header files, and function cross-references.

    \item \textit{Structure Initialization/Destruction Tool.} Identify potential init/cleanup functions for a given struct by scanning for functions that accept the structure pointer as an argument or return value.

    \item \textit{Code View Tool.} View code segments at a given file path and line number, allowing the LLM to drill down into specific regions.

    \item \textit{Driver Example Tool.} List existing driver files in the project to infer necessary header files.
\end{itemize}

\subsubsection{Implementation}

The seven tools above are implemented using two complementary backends: an LSP-based retriever for high semantic accuracy and a grammar tree-based retriever for robustness. HarnessAgent first attempts the LSP backend; if it fails, it falls back to the grammar tree parser.

\smallskip\noindent
\textbf{LSP-Based Retriever.}
We use \textit{clangd}~\cite{clangd} as the language server for C and C++. Given a symbol query, \textit{clangd} locates candidate positions across the project; we select the candidate with the largest namespace overlap to ensure accurate matching. Once located, we retrieve the symbol's definition, declaration, and cross-references via the LSP APIs.

To extract source code from the retrieved locations, we use \textit{Tree-Sitter}~\cite{treesitter} to parse the corresponding region. If parsing fails, we fall back to collecting a 50-line context window. Header files are inferred from definition/declaration locations by filtering for valid header extensions.

\smallskip\noindent
\textbf{Grammar Tree-Based Retriever.}
The LSP-based retriever may fail when (1) header files are auto-generated from templates, or (2) the codebase contains parsing/configuration errors that confuse \textit{clangd}.

As a fallback, we use a grammar tree-based retriever that directly parses source files with \textit{Tree-Sitter}. We first locate symbol occurrences via \textit{grep}, then traverse the grammar tree to extract declarations, definitions, and cross-references. This approach is more robust for incomplete or partially compilable projects.

Together, the two backends form a hybrid architecture: LSP provides precision when project metadata is available, while the grammar tree ensures resilience for noisy or broken codebases.

\subsection{Enhanced Validation}
\label{subsec:validation}

As identified in our study (\S\ref{subsec:limitation}), LLMs tend to generate fake or placeholder definitions to bypass validation checks such as successful compilation. This inflates apparent success rates while providing no actual fuzzing value.

To address this, we augment the standard validation pipeline with a \textit{fake-definition check}. Using \textit{Tree-Sitter}~\cite{treesitter}, we parse the generated harness and detect any locally defined functions sharing the same name as the target function. If such a definition exists, the harness is rejected, ensuring that it genuinely invokes the intended implementation rather than a fabricated stub.

\subsection{Compilation Error Triage}
\label{subsec:compilation}

As noted in our study, existing compilation workflows are brittle: they cannot distinguish harness-level errors from build-script issues, causing the LLM to generate misaligned fixes. To address this, we introduce a lightweight triage strategy that classifies errors and routes them appropriately.

We compile each harness by temporarily replacing an existing fuzz target in the project, avoiding modifications to build scripts. When compilation fails, we apply the following triage:

\smallskip\noindent
\textbf{Target switching.} We first retry with other existing fuzz targets in the project, as different targets may have different build configurations, thereby resolving the build issue.

\smallskip\noindent
\textbf{Link errors.} Undefined references typically indicate missing linkage rather than harness bugs. We retry with other targets; if all fail, we pass the error to the fixing module (e.g., to add missing \texttt{extern} declarations).

\smallskip\noindent
\textbf{Include path errors.} For errors like \lstinline[style=cbig]|'lwip/opt.h' file not found|, we parse the missing path from compiler logs and append the directory to \textit{CFLAGS}/\textit{CXXFLAGS}.

\smallskip\noindent
\textbf{Missing header dependencies.} When required headers are still missing, we provide existing driver examples to the fixing module so it can infer the necessary includes.

%% file: figure/studyframework.tex
\begin{figure*}[!t]
    \centering
   \includegraphics[width=\textwidth]{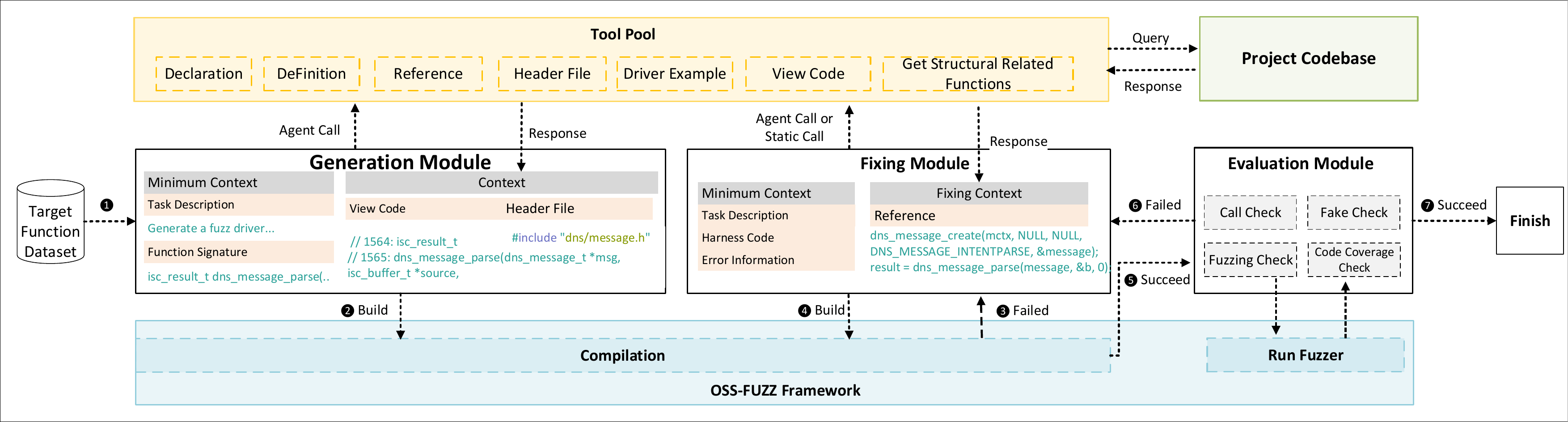}
    \caption{The agentic framework we enhanced for harness generation.}
    \label{fig:studyframework}
\end{figure*}

%% file: sec_05_evaluation.tex
\section{Evaluation of Enhancements}
This section evaluates whether the enhancements we proposed improve harness generation in practice.

\subsection{Experimental Setup}
All experiments were conducted on a desktop with a 12th Gen Intel(R) Core(TM) i9-12900K CPU, 64 GB RAM, and a 2 TB SSD. All methods are augmented with the fake check in the validation module by default.

\smallskip
\noindent\textbf{Baselines.} 
We apply our enhancements to the generation pipeline and compare against the three existing approaches discussed in~\S\ref{subsec:baselines}. We further introduce a baseline, \textit{Raw Model}, that uses the same framework as \our~while relying soly on error message feedback, without access to contextual information or tools. 

\smallskip
\noindent\textbf{Datasets.}
Rather than reusing the study benchmark from \S\ref{subsec:datasets}, we adopt the benchmark provided by \ossfuzzgen{} for two reasons: (1) the study benchmark was designed for detailed qualitative analysis and is too small for statistically meaningful evaluation, and (2) using cases we previously analyzed risks biasing results toward scenarios our enhancements were tuned to handle. This benchmark consists of 80 C projects and 256 C++ projects from the \ossfuzzgen~repository. We first exclude projects that fail to build either the Docker image or the fuzz targets. Next, we filter out functions that cannot be used to generate harnesses, including (1) functions defined within anonymous namespaces and (2) functions lacking explicit signatures. After filtering, we retain 65 C projects and 178 C++ projects. Each project contains up to five candidate functions; given the large workload, we evaluate one function per project. Because the candidate list is generated automatically and its ordering is unrelated to our method, choosing the first function yields an effectively random per-project sample, resulting in a final dataset of 243 target functions.

\smallskip
\noindent\textbf{Models.} 
We adopt \gptmini~ as our primary model for all experiments, as it provides an effective balance between model capability and monetary cost. 

\smallskip
\noindent\textbf{Evaluation Metric.} 
We adopt the same evaluation criteria from~\S\ref{subsec:criteria} and introduce additional quantitative metrics to comprehensively assess harness generation and fuzzing performance:

$\bullet$ \textit{SR@k.} A harness is considered successful if it passes the evaluation criteria. The success rate is defined as the proportion of successful functions out of the total number of evaluated functions, and serves as our primary performance metric. We further define \textit{SR@k} as the accumulated success rate over $k$ generation attempts.


$\bullet$ \textit{Target Function Branch Coverage.}
As overall harness coverage can be inflated by setup/teardown code and may not reflect whether the harness actually exercises the intended API, we measure the branch coverage at the \emph{target-function} level using the corpus generated during fuzzing. For each corpus input, we reset the coverage counters at the target function entry and record them again at function exit; the counter difference gives the target-function coverage for that input. We then aggregate this coverage across all inputs.


\input{table/overall}

\subsection{Performance of Harness Generation}

\autoref{table:overall} summarizes the harness generation performance. Across both C and C++ projects, \our{} achieves the highest \textit{SR@k} success rates. For C targets, \our~improves the \textit{SR@1} rate from 55.38\% (LLM4FDG) to 72.31\%, and for C++ targets from 51.68\% to 60.67\%. As the number of attempts increases, the gap widens further, reaching 81.53\% and 75.84\% at \textit{SR@3} for C and C++, respectively. These improvements are particularly notable for complex C++ projects where dependency resolution and namespace management are challenging. The consistent gains across \textit{SR@k} suggest that iterative retrieval and correction mechanisms effectively recover from initial failures and refine harnesses toward valid compilations.


To further analyze complementarity, \autoref{fig:overall_veen} presents the overlap of successful harnesses. Compared to \issta, \our\ adds 50 unique successes (vs. 7 unique to \issta), with 152 shared; compared to Raw, \our~adds 71 unique successes. The high overlap shows that our enhancements retain existing strengths, while the additional unique successes demonstrate extended capabilities. Since \sherpa~has few successful harnesses, we omit it in the remaining evaluation.

\input{figure/overall/overall_veen}

\input{table/ablation_component}

\subsubsection{Breakdown Analysis.}
To understand the contribution of each component (compilation, validation, and tool support), we disable them individually and measure the drop in success rate ($\Delta SR@3$) as shown in \autoref{table:ablation_component}. Disabling the validation component (Fake Check) leads to a far larger degradation (12.31\% for C and 17.98\% for C++), demonstrating that validation is critical for filtering incorrect or partially hallucinated harnesses. Compilation also has a noticeable impact: removing it decreases $\Delta SR@3$ by 6.2\% for C and 7.7\% for C++, indicating that fine-grained compilation error routing can mitigate the build issue to some extent.

Among the tooling modules, disabling any single tool consistently harms performance; for example, removing the header tool results in the largest drop (16.92\% and 21.35\%), followed by view code (13.84\% and 8.99\%). These results show that rich tool-assisted context retrieval is indispensable for function harness generation. Overall, the ablation indicates that all three components collectively reinforce the pipeline, and removing any of them substantially degrades performance.

\input{table/models}

\subsubsection{Impact of Model.}
To investigate the impact of models, we evaluate four representative LLMs: \claude, \gptmini, \qwen, and \ds.
\autoref{table:models} reports the harness-generation performance under both C and C++ settings. We measure performance using \textit{SR@1} as it sufficiently demonstrates the performance differences. Across all models and languages, \our~consistently outperforms the Raw Model, often by a large margin. Among those models, \gptmini~boosts the performance the most, 26\% for C and 21\% for C++.

Interestingly, the results suggest that tool usage ability matters more than pretrained knowledge for harness generation. For example, \qwen~outperforms \ds~and \gptmini~on the Raw model under C++, indicating stronger pretrained knowledge. However, with tool usage enabled, \ds~and \gptmini~surpass \qwen, demonstrating that effective tool use can compensate for weaker baseline performance.

\subsection{Quality of LLM-Generated Harness}
To assess the practical quality of the generated fuzz drivers, we fuzz each target for one hour with libfuzzer~\cite{libfuzzer} and categorize the outcomes into four groups:
\begin{itemize}
    \item \textit{Crash}: the harness triggers runtime failures, or address sanitizer violations~\cite{asan}.
    \item \textit{Not Reached}: the harness executes but never reaches the target function.
    \item \textit{Reached Without Coverage Increase}: the target function is reached, but its code coverage remains constant.
    \item \textit{Reached With Coverage Increase}: the target function's coverage increases during fuzzing.
\end{itemize}

\input{figure/cov_stat}

\smallskip
\noindent\textbf{Enhancements improve harness utility.}
As shown in \autoref{fig:cov-groupbar}, \our{} achieves the highest ratio of \emph{Reached With Coverage Increase} cases (75.0\%) and the lowest proportion of \emph{Not Reached} outcomes (12.5\%). Compared to the best-performing baseline (\issta{} at 69.2\% and 18.4\%, respectively), the enhancements reduce non-reachability by 6\% while increasing effective coverage increase cases by 6\%. 

Meanwhile, a small fraction of harnesses (5–10\%) triggered runtime crashes across all methods. After manual triage, we found that these crashes are caused by flawed harnesses (e.g., violated internal invariants) rather than bugs in the target functions. Importantly, \our{} maintains a comparable crash rate to the baselines, indicating that the improvements in reachability and coverage do not come at the cost of increased instability.

\input{figure/cov_distribution_pie}

\smallskip
\noindent\textbf{Enhancements improve code coverage.}
Beyond reachability, we analyze the target function coverage distribution (\autoref{fig:cov-distribution-pie}). 
HarnessAgent exhibits a lower proportion of low-coverage functions (1–25) and a slightly higher proportion of functions with coverage $\ge$ 100, with the overall distribution shifted toward higher coverage.

\input{table/tool_call_overall}

\subsection{Tool Effectiveness}
Our enhancements include tool-augmented context retrieval, enhanced validation, and compilation-error triage. The effectiveness of validation and compilation is reflected in \autoref{table:ablation_component}. However, this does not show whether the tools themselves return accurate responses. We therefore evaluate tool accuracy directly.

To evaluate efficiency and robustness, we log all tool calls (name/parameters) and their responses. We measure tool performance using the \textit{response rate} (non-empty responses / total calls), which may slightly overestimate accuracy but serves as a practical usability proxy. For comparison, we also query the local Fuzz Introspector~\cite{fuzz-introspector} under the same symbol requests and conditions; since it supports only header, declaration, definition, cross-reference, and view-code queries, we restrict the comparison to these five tools.

\smallskip
\noindent\textbf{Our tools are more effective.}
\autoref{table:tool_overall} shows our tool has the highest response rate across queries (94.29\% header, 93.91\% declaration, 93.79\% definition), exceeding LSP-only (83.21–88.44\%) and Fuzz Introspector (50.95–65.75\%). LSP accounts for most successes (over 80\%) by providing high-precision symbol localization; when it fails, we fall back to the parser. Although the parser alone has limited recall (no higher than 20\%), it resolves symbols that the LSP cannot handle, which improves robustness without sacrificing precision.

\input{figure/tool_call_project}

\smallskip
\noindent\textbf{Our tools are more robust.}
As shown in \autoref{fig:tool-dist}, most projects achieve response rates close to 100\%, indicating that our method performs reliably across diverse build environments and codebases. In contrast, Fuzz Introspector exhibits a broad and uneven distribution, with many projects clustered below 60\% or even failing entirely. This disparity highlights that our LSP–Parser hybrid design not only improves the average response rate but also significantly minimizes project-level failures, resulting in more consistent and robust performance across projects.

\section{The Road Ahead}
Our enhancements push harness generation closer to its practical limits, achieving over 80\% success rates with improved coverage and reachability. Yet two challenges remain before fully automated deployment: the cost of tool-augmented generation and distinguishing real bugs from false positives.

\input{table/efficiency}

\subsection{Efficiency and Cost}
\our{} improves quality but increases resource consumption. \autoref{table:efficiency} summarizes the total time, token usage, and monetary cost. \our{} consumes more tokens than other methods due to its extensive tool usage—each retrieval, validation, and correction step adds overhead. While the per-function cost remains manageable, scaling to large codebases with thousands of functions would amplify this cost substantially. Additionally, compilation remains a major contributor to overall running time, as \our{} recompiles frequently during iterative refinement. Reducing time and token consumption while maintaining generation quality remains an open challenge.

\subsection{Real-World Bug Discovery}
\input{table/targets}

To evaluate the practical security impact of \our, we applied it to 11 real-world projects, targeting the most recent release versions as shown in \autoref{table:targets}. We used an LLM-based scoring system to prioritize security-relevant functions (score $\ge 7$), capping at 300 functions per project. Each successfully compiled harness was fuzzed for one hour, as this duration provides a practical trade-off between exploration depth and computational cost.

\input{table/bugs}

\smallskip
\noindent\textbf{Bug discovery.}
As summarized in \autoref{table:bugs}, \our{} uncovered bugs spanning multiple classes, demonstrating that the generated harnesses can drive fuzzing deep enough to surface both high-impact memory corruptions and robustness failures in widely deployed libraries. Importantly, the discovered issues include bugs that have already been fixed upstream as well as vulnerabilities assigned with a CVE number, indicating that these are real bugs rather than harness artifacts.\looseness=-1

The identified bugs include multiple heap buffer overflows across several projects, notably \texttt{exiv2}, \texttt{lcms}, and \texttt{libssh}; most of these vulnerabilities have already been resolved by developers. Beyond memory corruption, \our{} also exposed resource-management failures, including a memory leak in \texttt{c-ares} and out-of-memory / timeout behaviors in \texttt{exiv2} and \texttt{libssh}.

\smallskip
\noindent\textbf{False positives remain a challenge.}
While \autoref{table:targets} reports 15 true positives, \our{} also triggered 43 false crashes, particularly in \texttt{libtiff} and \texttt{libpcap}. Unlike standard logical errors, these "false crashes" primarily stem from the semantic gap between automated harness generation and the strict internal invariants of complex C libraries. These failures fall into three main types:

\begin{itemize}[leftmargin=8pt, itemsep=2pt]
\item \textbf{State-machine violations:} Many crashes occurred when the harness invoked internal functions without satisfying the library's prerequisite state, triggering failed assertions or crashes.
\item \textbf{Resource-management mismatches:} The harness may free objects still owned by the library, miss required cleanup, or invoke the wrong free function, leading to double-free, use-after-free, out-of-memory, and memory leak.
\item \textbf{Argument mismatches:} The harness invokes internal APIs with semantically invalid arguments (e.g., incorrect buffer sizes, wrong enum values), causing crashes from invalid calling conventions.
\end{itemize}

\vspace{3pt}
These false positives highlight that LLM-generated harnesses still lack deep understanding of API semantics and ownership constraints. Reducing false positives remains an important direction for future work.

%% file: table/overall.tex
\begin{table}[t]
\centering
\footnotesize
\setlength\tabcolsep{1.9pt}
\caption{Harness generation performance. \textit{SR@k} denotes the
accumulated success rate over $k$ generation attempts. \label{table:overall}}

\begin{tabular}{llccccc}
\toprule
\multicolumn{2}{l}{\textbf{Success Rate}}    &\textbf{Raw Model} & \textbf{\ossfuzzgen} & \textbf{\issta}   & \textbf{Sherpa} & \textbf{\our} \\
\midrule
\multirow{2}{*}{SR@1} & C     & 46.15\%   & 46.15\%     & 55.38\% & 16.92\% & \textbf{72.31\%}    \\
                       & C++    & 39.32\%   & 38.76\%     & 51.68\% & 23.59\% & \textbf{60.67\%}    \\
 \midrule
\multirow{2}{*}{SR@2}& C      & 55.38\%   & 52.31\%     & 63.07\% & 18.46\% & \textbf{78.46\%}    \\
                       & C++   & 49.44\%   & 51.69\%     & 58.99\% & 24.16\% & \textbf{70.22\%}    \\
\midrule
\multirow{2}{*}{SR@3} & C     & 60.00\%   & 55.38\%     & 69.23\% & 18.46\% & \textbf{87.69\%}    \\
                       & C++    & 55.06\%   & 54.49\%     & 64.04\% & 24.72\% & \textbf{81.46\%}   \\
\bottomrule
\end{tabular}
\end{table}

%% file: figure/overall/overall_veen.tex
\begin{figure}[t!]
	\centering  
    \footnotesize
	\setlength\tabcolsep{2pt}
	\begin{tabular}{cccc}
             \includegraphics[width=0.11\textwidth]{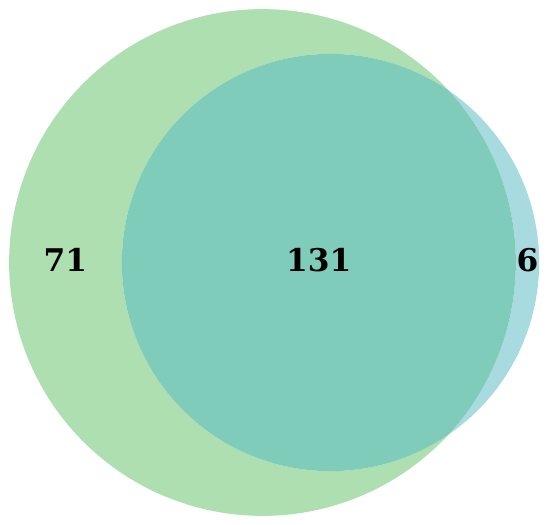} & 
              \includegraphics[width=0.11\textwidth]{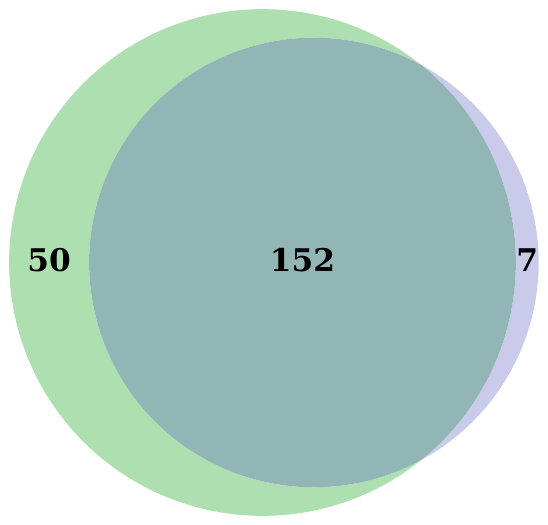}  &
              \includegraphics[width=0.11\textwidth]{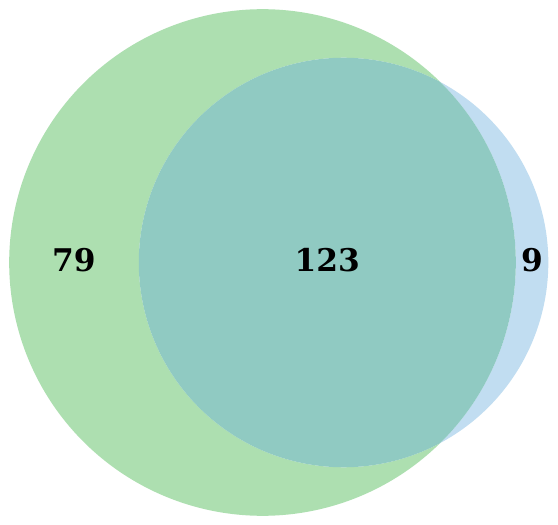} & 
              \includegraphics[width=0.11\textwidth]{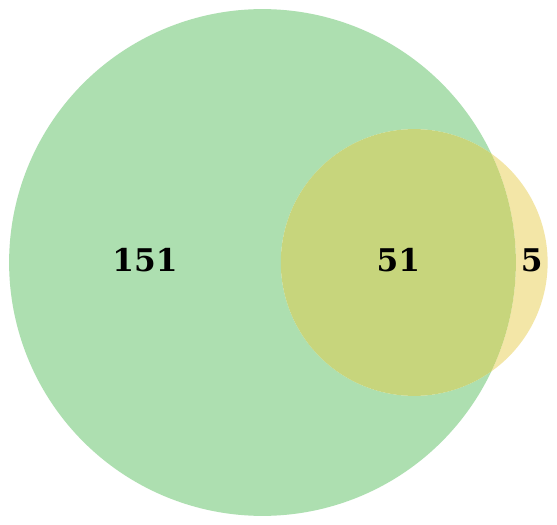} \\
              (a) Raw Model&(b) LLM4FDG&(c) OSS-Fuzz-Gen&(d) Sherpa 
	\end{tabular}
	\caption{Overlapped successful functions between HarnessAgent (left circle) and each evaluated method (right circle). Left and right values indicate functions uniquely generated by HarnessAgent and the compared method, respectively, while the center value indicates functions generated by both.}
\label{fig:overall_veen}
\end{figure}

%% file: table/ablation_component.tex
\begin{table}[t!]
\centering
\footnotesize
\caption{Impact of individual components on harness generation performance. Each value reports the change in success rate ($\Delta$SR@3) after disabling the corresponding component from the pipeline.\label{table:ablation_component}}
\vspace{-1em}
\renewcommand{\arraystretch}{1.2}
\setlength\tabcolsep{1pt}
\begin{tabular}{llcccccc}
\toprule
  &  &           &       & \multicolumn{4}{c}{\textbf{Tools}}\\
    \cmidrule(lr){5-8}
 \multirow{-2}{*}{\textbf{Disable}} &    & \multirow{-2}{*}{\makecell{\textbf{Compilation}}} & \multirow{-2}{*}{\makecell{\textbf{Fake} \\ \textbf{Check}}} & Header & Definition & Reference & View Code \\
\midrule
\textbf{C}         & $\Delta$SR@3(\%) &  6.2$\downarrow$  & 12.31$\downarrow$ & 16.92$\downarrow$  &4.62$\downarrow$  &7.69 $\downarrow$  & 13.84$\downarrow$         \\
\textbf{C++ }      & $\Delta$SR@3(\%) & 7.7$\downarrow$  & 17.98$\downarrow$   & 21.35$\downarrow$  &6.18$\downarrow$ &7.30  $\downarrow$ & 8.99$\downarrow$      \\
\bottomrule
\end{tabular}
\end{table}

%% file: table/models.tex
\begin{table}[t]
    \centering
    \footnotesize
    \tabcolsep=5.9pt

    \caption{Harness generation performance with different models. The values represent the \textit{SR@1}(\%) success rates. \label{table:models}}
    \vspace{-1em}
    \begin{tabular}{llcccc}
    \toprule
    \multicolumn{2}{l}{\textbf{Method}}   & \textbf{\claude} & \textbf{\gptmini} & \textbf{\qwen}  & \textbf{\ds}    \\
    \midrule
    \multirow{2}{*}{\textbf{C}}   & Raw Model   & 38.46 & 46.15 & 36.92 & 40.00 \\
                         & \our & 63.08  & 72.31 & 46.15 & 49.23 \\
                         \midrule
    \multirow{2}{*}{\textbf{C++}} & Raw Model  & 42.13  & 39.32 & 40.45 & 34.83 \\
                         & \our & 64.61  & 60.67 & 48.31 & 51.12 \\
                         \bottomrule
    \end{tabular}
\end{table}

%% file: figure/cov_stat.tex
\begin{figure}[t!]
    \centering
   \includegraphics[width=\columnwidth]{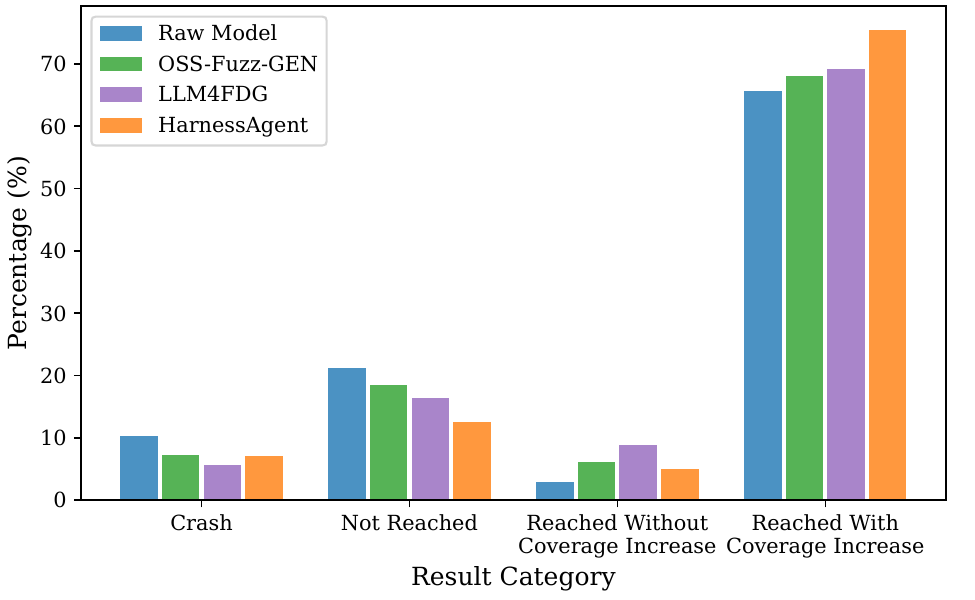}
    \vspace{-2em}
    \caption{One-hour fuzzing results of evaluated methods.}
    \label{fig:cov-groupbar}
\end{figure}


%% file: figure/cov_distribution_pie.tex
\begin{figure}[t!]
    \centering
   \includegraphics[width=0.98\columnwidth]{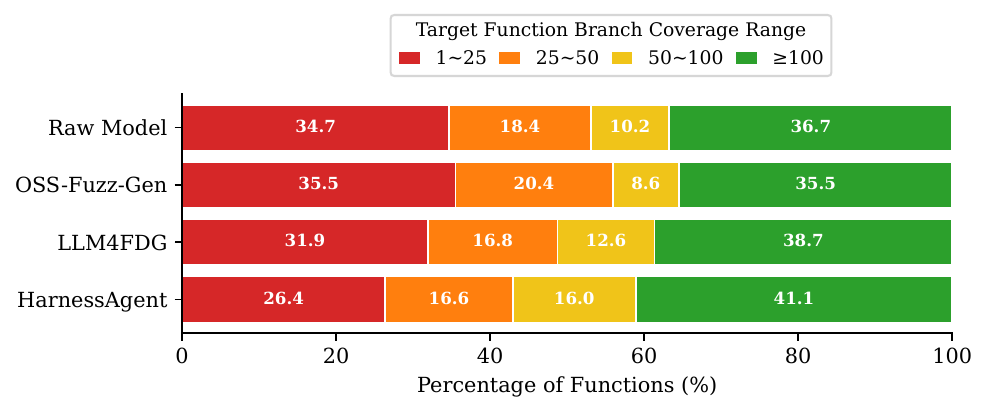}
    \vspace{-1em}
    \caption{Distribution of functions across branch coverage ranges for each method. Values in the color stack denote the percentage of functions in each branch coverage range.}
    \label{fig:cov-distribution-pie}
\end{figure}

%% file: table/tool_call_overall.tex
\begin{table}[t!]
\centering
\footnotesize
\tabcolsep=2.2pt
\caption{Comparison of tool call response rate (\%). \label{table:tool_overall}}
\vspace{-1.0em}
\begin{tabular}{lccccc}
\toprule
\textbf{Tool} & \textbf{Header} & \textbf{Declaration} & \textbf{Definition} & \textbf{Reference} & \textbf{View code}  \\
\toprule
\textbf{Introspector}  & 50.95  & 58.28       & 65.75      & 19.42     & 94.85                \\
\midrule
\textbf{\our}           & \textbf{94.29}  & \textbf{93.91}       & \textbf{93.79}      & \textbf{66.54}     & \textbf{96.70}             \\
\texttt{- LSP}           & 83.21  & 85.16       & 88.44      & 47.08       & -                 \\
\texttt{- Parser}        & 11.07  & 8.75        & 5.35       & 19.46     & -                \\
\midrule
\textbf{\#Total Call}       & 840    & 640         & 692        & 484       & 1941               \\
\bottomrule
\end{tabular}
\end{table}


%% file: figure/tool_call_project.tex
\begin{figure}[t]
    \centering
   \includegraphics[width=0.95\columnwidth]{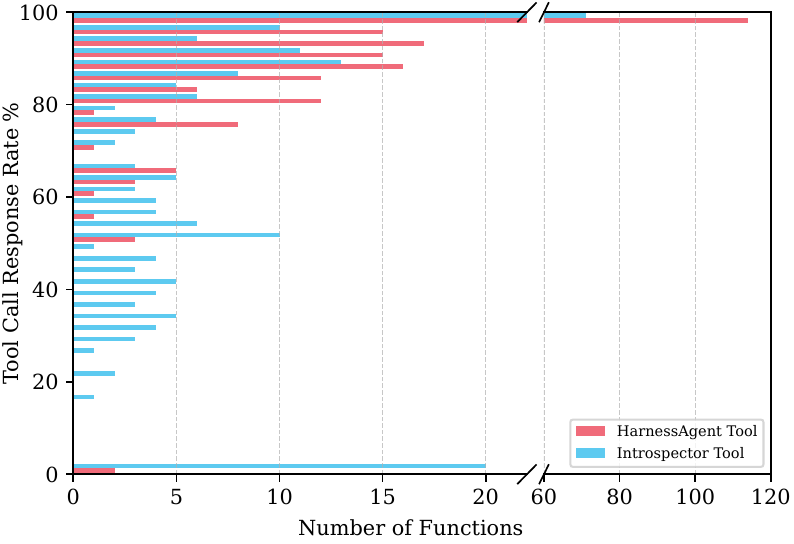}
    \vspace{-1em}
   \caption{Tool response rate distribution across target functions.}
    \label{fig:tool-dist}
\end{figure}


%% file: table/efficiency.tex
\begin{table}[t!]
    \centering
    \tabcolsep=7.5pt
    \caption{Time consumption, token usage, and monetary cost of different methods. \texttt{Avg.} means the average cost computed over the successful functions.}
    \label{table:efficiency}
    \vspace{-5pt}
    \footnotesize
    \begin{tabular}{lccccc}
    \toprule
    \multicolumn{1}{l}{\multirow{2}{*}{\textbf{Method}}} & \multirow{2}{*}{\textbf{Total Time}} & \multicolumn{2}{c}{\textbf{Total Token(M)}} & \multicolumn{2}{c}{\textbf{Cost(\$)}} \\
    \cmidrule(lr){3-4} \cmidrule(lr){5-6}
                          &               & Input          & Output         & Total       & Avg.        \\
    \midrule
    \textbf{Raw Model}                                   & 11h08m                 & 4.815          & 6.494          & 13.78       & 0.138      \\
    \textbf{\ossfuzzgen}                                 &04h18m                 & 2.075          & 3.000          & 6.460       & 0.057      \\
    \textbf{\issta}                                      &08h12m                 & 6.839          & 4.341          & 9.855       & 0.077      \\
    \textbf{\our}                                        &09h17m                 & 65.186         & 11.869         & 35.73       & 0.215     \\
    \bottomrule
    
    \end{tabular}
\end{table}

    

%% file: table/targets.tex
\begin{table}[t]
\centering
\footnotesize
\setlength\tabcolsep{3.2pt}
\renewcommand{\arraystretch}{1.1}

\caption{Statistics of \our-generated fuzz harnesses. ``Total'' means the total number of functions identified via \texttt{clangd}; ``Selected'' represents the functions selected for fuzzing by the LLM (score $\ge 7$, top 300); ``SR'' refers to the functions for which \our{} successfully generates a fuzz harness.\label{table:targets}}
\vspace{-5pt}
\begin{tabular}{lccccccc}
\toprule
\multicolumn{1}{l}{\multirow{2}{*}{\textbf{Library}}} &
\multicolumn{1}{l}{\multirow{2}{*}{\textbf{Commit ID}}} &
\multicolumn{1}{l}{\multirow{2}{*}{\textbf{Version}}} &
\multicolumn{3}{c}{\textbf{\# Function}} &
\multicolumn{2}{c}{\textbf{Result}} \\
\cmidrule(lr){4-6} \cmidrule(lr){7-8}
& & & \texttt{Total} & \texttt{Selected} & \texttt{SR} & \texttt{\#TP} & \texttt{\#FP} \\
\midrule
c-ares        & 3ac47ee   & 1.34.6     & 952           & 118            & 103       & 1    & 1    \\
cjson         & c859b25   & 1.7.19     & 150           & 36             & 35        & 0    & 0    \\
exiv2         & afcb7a8   & 0.28.7     & 2971          & 172            & 143       & 3    & 3    \\
lcms          & 35b57af   & 2.18       & 1230          & 241            & 231       & 2    & 4    \\
libaom        & d772e33   & 3.13.1     & 6017          & 300            & 218          & 0    & 1    \\
libjpeg-turbo & af9c1c2   & 3.1.3      & 1288          & 300            & 166       & 1    & 3    \\
liblouis      & b798a79   & 3.36.0     & 508           & 137            & 130       & 4    & 1    \\
libpcap       & a999701   & 1.10.      & 561           & 81             & 73        & 0    & 9    \\
libssh        & 301d0e1   & 0.11.3     & 1333          & 300            & 283       & 3    & 6    \\
libtiff       & 5fe20d0  & 4.7.1      & 1311          & 276            & 229       & 1    & 11   \\
libvpx        & 1024874   & 1.16.0-rc1 & 2228          & 300            & 243          & 0    & 4   \\
\midrule
Total  & -   & - & -          & -            &   -       & 15    & 43  \\
\bottomrule
\end{tabular}
\end{table}

%% file: table/bugs.tex
\begin{table}[t]
\footnotesize
\tabcolsep=8pt
\caption{New bugs found by \our\label{table:bugs}}
\vspace{-5pt}
\begin{tabular}{llll}
\toprule
\textbf{Library}          & \textbf{ID}                  & \textbf{Bug Type}             & \textbf{Status}    \\
\midrule
c-ares                    & Issue 1075          & Memory Leak          & Fixed  \\
\midrule
\multirow{3}{*}{exiv2}    & Issue 3468  & Heap Buffer Overflow & Fixed  \\
                          & Issue 3454          & Out of Memory        & Fixed  \\
                          & CVE-2026-27596  & Heap Buffer Overflow & Fixed  \\
\midrule
\multirow{2}{*}{lcms} & Commit da6110b      & Heap Buffer Overflow  & Fixed     \\
 & Commit 6a68601     & Heap Buffer Overflow  & Fixed     \\
 \midrule
libjpeg-turbo             & Commit 9131c06      & Heap Buffer Overflow & Fixed     \\
\midrule
\multirow{4}{*}{liblouis} & Issue 1904          & Heap Buffer Overflow & Fixed \\
                          & Issue 1910          & Infinite Loop & Reported          \\
                          & Issue 1910          & Infinite Loop & Reported          \\
                          & Issue 1910          & Infinite Loop & Reported          \\      
\midrule
\multirow{3}{*}{libssh}   & CVE Asigned         & Heap Buffer Overflow & Fixed \\
                          & CVE Asigned         & Infinite Loop   & Fixed \\
                          & CVE Asigned         & Timtout              & Fixed \\
\midrule
libtiff                   & Issue 780           & Heap Buffer Overflow & Reported \\
\bottomrule
\end{tabular}
\end{table}

%% file: sec_06_relatedwork.tex
\section{Related Work}



\subsection{Automatic Fuzzing Harness Generation}

Existing efforts for automatic fuzzing harness generation can be broadly categorized into \textit{static} and \textit{dynamic} approaches, depending on whether they rely on code analysis or runtime feedback to construct valid API call sequences.

Static approaches infer API dependencies and argument structures from reference code. FUDGE~\cite{FUDGE} slices consumer programs to recover usage contexts, while FuzzGen~\cite{fuzzgen}, Winnie~\cite{jung2021winnie}, and APICraft~\cite{apicraft} build dependency graphs/traces to reconstruct API relations; RULF~\cite{jiang2021rulf} and RPG~\cite{xu2024rpg} synthesize API sequences via type and dependency reasoning. These methods work well for API-level targets but rely on analyzable consumer code or clear specifications, which are often missing for internal or highly stateful components.

Dynamic approaches construct harnesses via runtime observation and feedback. Hopper~\cite{hopper} composes API sequences using interpreter execution/coverage feedback, and OGHarn~\cite{capiharness} mutationally stitches harnesses from headers and keeps those passing compilation/execution/coverage oracles. GraphFuzz~\cite{green2022graphfuzz} and AFGen~\cite{afgen} perform synthesis during fuzzing using coverage and learned dependencies, while NEXZZER~\cite{NEXZZER} updates dynamic API graphs with fuzzing feedback and type matching; Daisy~\cite{zhangdaisy} mines and merges object-usage sequences from real consumers. These methods reduce dependence on consumer code but still rely on schemas or manual/domain constraints to maintain correctness.

Overall, static and dynamic methods improve automation but still struggle with semantic constraints and adaptability, often yielding false positives; this motivates LLM-based approaches that infer richer context and usage constraints.

\subsection{LLM for Fuzzing Harness Generation}

Recent work uses LLMs to synthesize fuzzing harnesses by inferring API usage constraints from code and lightweight program context. Existing approaches fall into two categories: API harness generation and target function harness generation.

\smallskip\noindent
\textbf{API Harness Generation.}
Several prior works~\cite{promptfuzzingccs24,promefuzzccs25,ckgfuzzer} leverage large language models to generate fuzzing harnesses for API functions. Rather than targeting a specific API, these approaches construct harnesses from a collection of APIs and rely on the LLM to determine which functions to invoke. Representative systems include PromptFuzz~\cite{promptfuzzingccs24}, PromeFuzz~\cite{promefuzzccs25}, and CKGFuzzer~\cite{ckgfuzzer}, which primarily differ in the amount and type of contextual information they provide. PromptFuzz uses API signatures and documentation, CKGFuzzer augments prompts with call-graph knowledge and refines generated drivers and seeds, and PromeFuzz integrates multiple graph representations along with a sanitizer module for refinement and crash triage. While effective, these approaches are largely limited to library projects and API functions, restricting their broader applicability.

\smallskip\noindent
\textbf{Target Function Harness Generation.} Instead of focusing solely on API functions, this line of work generates a harness for arbitrary functions. 
\ossfuzzgen~\cite{OSS-Fuzz-Gen} iteratively generates and compiles harnesses with rule-based feedback from the build/fuzzing pipeline. \issta~\cite{issta} studies how different prompts and fixing strategies affect the harness generation. \sherpa~\cite{sherpa2024} uses a general-purpose code agent (Codex~\cite{codex}) to reduce pipeline engineering, but still often struggles with compilation. 


%% file: sec_07_discussion.tex




%% file: sec_09_conclusion.tex
\section{Conclusion}
This paper presents a study of LLM-based harness generation for fuzzing, identifying key limitations in how existing methods provision context and validate outputs. Our analysis reveals that static context retrieval and shallow validation are primary causes of generation failures. To address these issues, we introduce enhancements including tool-augmented context retrieval, strengthened validation with fake-definition detection, and compilation-error triage. Evaluation on 243 OSS-Fuzz targets shows that these enhancements improve the three-shot success rate by approximately 20\% over existing techniques. However, challenges remain: tool-augmented generation incurs a higher cost, and LLM-generated harnesses still produce false positives due to limited understanding of API semantics and ownership constraints.

\section*{Data Availability Statement}
All our source code and datasets are anonymized and released for review at \url{https://doi.org/10.5281/zenodo.19337585}.